\documentclass[11pt]{article}
\usepackage{epsf,cite}

\def\bq{\begin{eqnarray}}
\def\eq{\end{eqnarray}}
\def\be{\begin{eqnarray}}
\def\ee{\end{eqnarray}}
\def\ben{\begin{enumerate}}\def\een{\end{enumerate}}

\def\roughly#1{\mathrel{\raise.3ex\hbox{$#1$\kern-.75em
\lower1ex\hbox{$\sim$}}}}

\def\bra{\langle }
\def\ket{\rangle }

\begin{document}

\title{Generalized parton distributions and the parton structure 
of light nuclei}

\author{Sergio Scopetta\thanks{Invited talk at Light Cone 2010: 
Relativistic Hadronic and Particle Physics,
June 14-18, 2010,
Valencia, Spain. To appear in PoS, Proceedings of Science}
\\ University of Perugia, Italy\\
E-mail: sergio.scopetta@pg.infn.it}
\maketitle
\begin{abstract}
The measurement of nuclear Generalized Parton Distributions (GPDs)
represents a valuable tool to understand the structure of bound
nucleons and the phenomenology of hard scattering off nuclei.
By using a realistic, non-relativistic microscopic approach 
for the evaluation of GPDs of
$^3$He,
it will be shown that conventional nuclear effects, such as
isospin and binding ones, or the uncertainty related to the use of
a given nucleon-nucleon potential, are bigger than in the
forward case so that, if great attention is not
paid, conventional nuclear effects can be easily
mistaken for exotic ones. It is stressed that $^3$He,
for which the best realistic calculations are possible,
represents a unique target to discriminate between conventional
and exotic effects. The complementary information which
could be obtained by using a $^3$H target,
the possible extraction of the neutron information,
as well as the relevance of a relativistic treatment,
will be also addressed.
\end{abstract}

\section{Introduction}

The measurement of
Generalized Parton Distributions (GPDs) \cite{first}, 
parametrizing  the non-perturbative hadron structure
in hard exclusive 
processes,
represents one of the challenges of nowadays hadronic Physics.
GPDs enter
the long-distance dominated part of
exclusive lepton Deep Inelastic Scattering
(DIS) off hadrons.
Deeply Virtual Compton Scattering (DVCS),
i.e. the process
$
e H \longrightarrow e' H' \gamma
$ when
$Q^2 \gg m_H^2$,
is one of the the most promising to access GPDs
(here and in the following,
$Q^2$ is the momentum transfer between the leptons $e$ and $e'$,
and $\Delta^2$ the one between the hadrons $H$ and $H'$)
\cite{first}.
Relevant experimental efforts to measure GPDs
are taking place, and
a few DVCS data have been already published 
\cite{hermes}. 
The issue of measuring GPDs for nuclei
has been addressed in several papers
\cite{cano1}.
While some studies have shown that the measurement
of nuclear GPDs can unveil information on possible
medium modifications of nucleons in nuclei \cite{liuti2}, 
great attention has to be paid to avoid to mistake them with
conventional nuclear effects.
To this respect, a special role
would be played by few body nuclear targets,
for which realistic studies 
are possible and exotic effects, such as
the ones of non-nucleonic degrees of freedom,
not included in a
realistic wave function,
can be disentangled.
To this aim,
in Ref. \cite{io}, a realistic IA calculation
of the quark unpolarized GPD $H_q^3$ of
$ ^3He $ has been presented.
The study of GPDs for $^3He$ is interesting
for many aspects. 
In fact, $^3He$ is a well known nucleus, 
and it is extensively used as an effective neutron target:
the properties of the free neutron are being investigated
through experiments with nuclei, whose data
are analyzed taking nuclear effects properly into account.
For example, it has been shown, firstly in \cite{thom1},
that unpolarized DIS off trinucleons ($^3H$ and $^3He$) can provide 
relevant information on PDFs at large $x_{Bj}$, 
while it is known since a long time that 
its particular spin structure
suggests the use of $^3He$ as an effective polarized
neutron target \cite{friar}.
Polarized $^3He$ will be therefore the first candidate
for experiments aimed at the study of spin-dependent
GPDs of the free neutron. 
In Ref. \cite{io}, the GPD $H_q^3$ of
$ ^3He $ has been evaluated
using a realistic non-diagonal spectral function,
so that momentum and binding effects are rigorously estimated.
The scheme proposed in that paper is valid for $\Delta^2 \ll Q^2,M^2$
and it permits to calculate GPDs in the kinematical range relevant to
the coherent, no break-up channel of deep exclusive processes off $^3He$.
In fact, the latter channel can be hardly studied at
large $\Delta^2$, due to the vanishing cross section.
Nuclear effects are found to be larger than in the forward case
and to increase with $\Delta^2$ at fixed
skewedness, and with the skewedness at fixed $\Delta^2$.
In particular the latter $\Delta^2$ dependence
does not simply factorize,
in agreement with previous findings for the deuteron target
and at variance
with prescriptions proposed for finite nuclei. 

Here, the analysis of Ref. \cite{iol}, which extended 
that of Ref. \cite{io} into various directions, is reviewed.
The main point of the contribution will be to
stress that the properties of nuclear GPDs should not be trivially
inferred from those of nuclear parton distributions. 
\section{Conventional nuclear effects on the GPDs of $^3He$}
Let us introduce the definition GPDs 
to be used in what follows.
For a spin $1/2$ hadron target, with initial (final)
momentum and helicity $P(P')$ and $s(s')$, 
respectively, 
the GPDs $H_q(x,\xi,\Delta^2)$ and
$E_q(x,\xi,\Delta^2)$
are defined through the light cone correlator
\begin{eqnarray}
\label{eq1}
F^q_{s's}(x,\xi,\Delta^2) & = &
{1 \over 2} \int {d \lambda \over 2 \pi} e^{i \lambda x}
\bra P' s' | \, \bar \psi_q \left(- {\lambda n \over 2}\right)
\slash{n} \, \psi_q \left({\lambda n \over 2} \right) | P s \ket    
\nonumber \\
& = & H_q(x,\xi,\Delta^2) {1 \over 2 }\bar U(P',s') 
\slash{n} U(P,s) 
\nonumber \\
& + & 
E_q(x,\xi,\Delta^2) {1 \over 2} \bar U(P',s') 
{i \sigma^{\mu \nu} n_\mu \Delta_\nu \over 2M} U(P,s)~,
\nonumber
\end{eqnarray}
where 
$\Delta=P^\prime -P$
is the 4-momentum transfer to the hadron,
$\psi_q$ is the quark field and M is the hadron mass.
It is convenient to work in
a system of coordinates where
the photon 4-momentum, $q^\mu=(q_0,\vec q)$, and $\bar P=(P+P')/2$ 
are collinear along $z$.
The skewedness variable, $\xi$, is defined as
\bq
\xi = - {n \cdot \Delta \over 2} = - {\Delta^+ \over 2 \bar P^+}
= { x_{Bj} \over 2 - x_{Bj} } + {\cal{O}} \left ( {\Delta^2 \over Q^2}
\right ) ~,
\label{xidef}
\eq
where $n$
is a light-like 4-vector
satisfying the condition $n \cdot \bar P = 1$.
(Here and in the following, $a^{\pm}=(a^0 \pm a^3)/\sqrt{2}$).
In addition to the variables
$x,\xi$ and $\Delta^2$, GPDs depend
on the momentum scale $Q^2$. 
Such a dependence, not discussed here, will be 
omitted.
The constraints of $H_q(x,\xi,\Delta^2)$ are: 
i) the 
``forward'' limit, 
$P^\prime=P$, i.e., $\Delta^2=\xi=0$, yielding the usual PDFs
\bq
H_q(x,0,0)=q(x)~;
\label{i)}
\eq
ii)
the integration over $x$, yielding the contribution
of the quark of flavour $q$ to the Dirac 
form factor (f.f.) of the target:
\bq
\int dx H_q(x,\xi,\Delta^2) = F_1^q(\Delta^2)~;
\label{ii)}
\eq
iii) the polynomiality property,
involving higher moments of GPDs.
In Ref. \cite{epj},
an expression
for $H_q(x,\xi,\Delta^2)$ of a given hadron target, 
for small values of $\xi^2$, has been obtained
from the definition Eq. (\ref{eq1}).
The approach has been later applied 
in Ref. \cite{io} to obtain the GPD 
$H_q^3$ of $^3He$ in IA, as
a convolution between
the non-diagonal spectral function of the internal nucleons,
and the GPD $H_q^N$ of the nucleons themselves. 
Let me recall the main formalism of Ref. \cite{io},
which will be used in this paper.
In the class of frames discussed above,
and in addition to the kinematical variables
$x$ and $\xi$, already defined,
one needs the corresponding ones for the nucleons in the target nuclei,
$x'$ and $\xi'$. 
The latter quantities can be obtained defining the ``+''
components of the momentum $k$ and $k + \Delta$ of the struck parton
before and after the interaction, with respect to
$\bar P^+$ and $\bar p^+ = {1 \over 2} (p + p')^+$ (see \cite{io}
for details).
In Ref. \cite{io},
a convolution formula for $H_q^3$ has been derived in IA,
using the standard procedure developed in studies of
DIS off nuclei \cite{fs,cio,ia}.
It reads:
\begin{eqnarray}
H_q^3(x,\xi,\Delta^2) & \simeq & 
\sum_N \int dE \int d \vec p
\, 
[ P_{N}^3(\vec p, \vec p + \vec \Delta, E ) + 
{\cal{O}} 
( {\vec p^2 / M^2},{\vec \Delta^2 / M^2}) ]
\nonumber
\\
& \times & 
{\xi' \over \xi}
H_{q}^N(x',\xi',\Delta^2) + 
{\cal{O}} 
\left ( \xi^2 \right )~.
\label{spec}
\end{eqnarray}
In the above equation, 
$P_{N}^3 (\vec p, \vec p + \vec \Delta, E )$ is
the one-body non-diagonal spectral function
for the nucleon $N$,
with initial and final momenta
$\vec p$ and $\vec p + \vec \Delta$, respectively,
in $^3He$:
\begin{eqnarray}
P_N^3(\vec p, \vec p + \vec \Delta, E)  & = & 
{1 \over (2 \pi)^3} {1 \over 2} \sum_M 
\sum_{R,s}
\bra \vec P'M | (\vec P - \vec p) S_R, (\vec p + \vec \Delta) s\ket 
\nonumber
\\
& \times & 
\bra (\vec P - \vec p) S_R,  \vec p s| \vec P M \ket
\, \delta(E - E_{min} - E^*_R)~,
\label{spectral}
\end{eqnarray}
and the quantity $H_q^N(x',\xi',\Delta^2)$
is
the GPD of the bound nucleon N
up to terms of order $O(\xi^2)$.
The delta function in Eq (\ref{spectral})
defines $E$, the removal energy, in terms of
$E_{min}=| E_{^3He}| - | E_{^2H}| = 5.5$ MeV and
$E^*_R$, the excitation energy 
of the two-body recoiling system.
The main quantity appearing in the definition
Eq. (\ref{spectral}) is
the overlap integral
\bq
\bra \vec P M | \vec P_R S_R, \vec p s \ket =
\int d \vec y \, e^{i \vec p \cdot \vec y}
\bra \chi^{s},
\Psi_R^{S_R}(\vec x) | \Psi_3^M(\vec x, \vec y) \ket~,
\label{trueover}
\eq 
between the eigenfunction 
$\Psi_3^M$ 
of the ground state
of $^3He$, with eigenvalue $E_{^3He}$ and third component of
the total angular momentum $M$, and the
eigenfunction $\Psi_R^{S_R}$, with eigenvalue
$E = E_{min}+E_R^*$ of the state $R$ of the intrinsic
Hamiltonian pertaining to the system of two interacting
nucleons.
As discussed in Ref. \cite{io}, the accuracy of the calculations
which will be presented, since a NR spectral function
will be used to evaluate Eq. (\ref{spec}), is
of order 
${\cal{O}} 
\left ( {\vec p^2 / M^2},{\vec \Delta^2 / M^2} \right )$,
or, which is the same,
$\vec p^2, \vec \Delta^2 << M^2$.
The interest of the present
calculation is indeed to investigate nuclear effects at low values
of $\vec \Delta^2$, for which measurements in the coherent channel 
may be performed.
The main emphasis of the present approach, as already said,
is not on the absolute values of the results, but in the nuclear effects,
which can be estimated by taking any reasonable form for
the internal GPD.
Eq. (\ref{spec}) can be written in the form
\begin{eqnarray}
H_{q}^3(x,\xi,\Delta^2) =  
\sum_N \int_x^1 { dz \over z}
h_N^3(z, \xi ,\Delta^2 ) 
H_q^N \left( {x \over z},
{\xi \over z},\Delta^2 \right)~,
\label{main}
\end{eqnarray}
where the off-diagonal light cone momentum distribution
\begin{equation}
h_N^3(z, \xi ,\Delta^2 ) =  
\int d E
\int d \vec p
\, P_N^3(\vec p, \vec p + \vec \Delta) 
\delta \left( z + \xi  - { p^+ \over \bar P^+ } \right)
\label{hq0}
\end{equation}
has been introduced.
As it is shown in Ref. \cite{io}, 
Eqs. (\ref{main}) and (\ref{hq0}) or, which is the same,
Eq. (\ref{spec}), fulfill the constraint $i)-iii)$ previously listed.
The constraint $i)$, i.e. the forward limit
of GPDs, is verified
by taking
the forward limit ($\Delta^2 \rightarrow 0, \xi \rightarrow 0$)
of Eq. (\ref{main}), 
yielding the parton distribution $q_3(x)$ in IA:
\cite{fs,cio,io2}: 
\begin{eqnarray}
q_3(x) =  H_q^3(x,0,0) =
\sum_{N} \int_x^1 { dz \over z}
f_{N}^3(z) \,
q_{N}\left( {x \over z}\right)~.
\label{mainf}
\end{eqnarray}
In the latter equation,
\begin{equation}
f_{N}^3(z) = h_{N}^3(z, 0 ,0) =  \int d E \int d \vec p
\, P_{N}^3(\vec p,E) 
\delta\left( z - { p^+ \over \bar P^+ } \right)
\label{hq0f}
\end{equation}
is the forward limit of Eq. (\ref{hq0}), i.e.
the light cone momentum distribution of the nucleon $N$
in the nucleus, $q_N(x)= H_q^N( x , 0, 0)$
is the distribution
of the quark of flavour $q$ 
in the nucleon $N$ and $P_N^3(\vec p, E)$,
the $\Delta^2 \longrightarrow 0$ limit of
Eq. (\ref{main}), is the
one body spectral function.
The constraint $ii)$, i.e. the $x-$integral of the GPD
$H_q$, is also 
fulfilled. By $x-$integrating Eq. (\ref{main}),
one obtains the contribution, 
of the quark of flavour $q$,
to the
nuclear f.f.
Eventually the polynomiality, condition $iii)$,
is formally fulfilled by Eq. (\ref{spec}).

In the following,
$H_q^3(x,\xi,\Delta^2)$, Eq. (\ref{spec}), 
will be evaluated in the nuclear Breit Frame.
The non-diagonal spectral function
Eq. (\ref{spectral}), appearing in Eq.
(\ref{spec}),
will be calculated 
by means of 
the overlap Eq. (\ref{trueover}), which 
exactly includes
the final state interactions in the two nucleon recoiling system
The realistic wave functions $\Psi_3^M$
and $\Psi_R^{S_R}$ in Eq. (\ref{trueover})
have been evaluated
using the 
AV18 interaction \cite{av18}.
In particular $\Psi_3^M$ has been 
developed along the lines of Ref. \cite{tre}.
The same overlaps, evaluated along the line
of Ref. \cite{gema}, have been already used in Ref.\cite{io,io2}.

The other ingredient in Eq. (\ref{spec}), i.e.
the nucleon GPD $H_q^N$, has been modelled in agreement with
the Double Distribution representation \cite{radd}, as described
in \cite{rad1} (See Ref. \cite{io}).
In Ref. \cite{io} it has been shown that the described formalism reproduces 
well, in the proper limits, the IA results for nuclear parton distributions
and form factor. In particular, in the latter case, the IA calculation
reproduces well
the data up to a momentum transfer $-\Delta^2=0.25$ GeV$^2$,
which is enough for the aim of this calculation.
In fact, the region of higher momentum transfer 
is not considered here, being
phenomenologically not relevant for the calculation
of GPDs entering coherent processes.

Conventional nuclear effects on the GPDs
of $^3He$ will be now discussed. The aim is that of avoiding to mistake
them for exotic ones in possible measurements of nuclear GPDs,
and to stress the relevance of experiments using $^3He$ targets.
As already done in Ref. \cite{io},
the full result for $H_q^3$, Eq. (\ref{spec}),
will be 
compared with a prescription
based on the assumptions
that nuclear effects are neglected
and the global $\Delta^2$ dependence is
described 
by
the f.f. of $^3He$:
\bq
H_q^{3,(0)}(x,\xi,\Delta^2) 
= 2 H_q^{3,p}(x,\xi,\Delta^2) + H_q^{3,n}(x,\xi,\Delta^2)~,
\label{app0}
\eq
where the quantity
\bq
H_q^{3,N}(x,\xi,\Delta^2)=  
\tilde H_q^N(x,\xi)
F_q^3 (\Delta^2)
\label{barh}
\eq
represents effectively the flavor $q$ GPD of the bound nucleon 
$N=n,p$ in $^3He$. Its $x$ and $\xi$ dependences, given by 
$\tilde H_q^N(x,\xi)$, 
are the same of the GPD of the free nucleon $N$,
while its $\Delta^2$ dependence is governed by the
contribution of the flavor $q$ to the
$^3He$ f.f., $F_q^3(\Delta^2)$.
The effect of nucleon motion
and binding can be shown through 
the ratio
\be
R_q(x,\xi,\Delta^2) = { H_q^3(x,\xi,\Delta^2) \over H_q^{3,(0)}
(x,\xi,\Delta^2)}~, 
\label{rnew}
\eq
i.e. the ratio
of the full result, Eq. (\ref{spec}),
to the approximation Eq. (\ref{app0}).
The ratio Eq. (\ref{rnew})
shows nuclear effects in a very natural way.
As a matter of facts, its forward limit 
yields an
EMC-like ratio for the parton distribution $q$ and,
if $^3He$ were made of free nucleons at rest, it would be one.

\begin{figure}[ht]
\vspace{10.0cm} 
\includegraphics{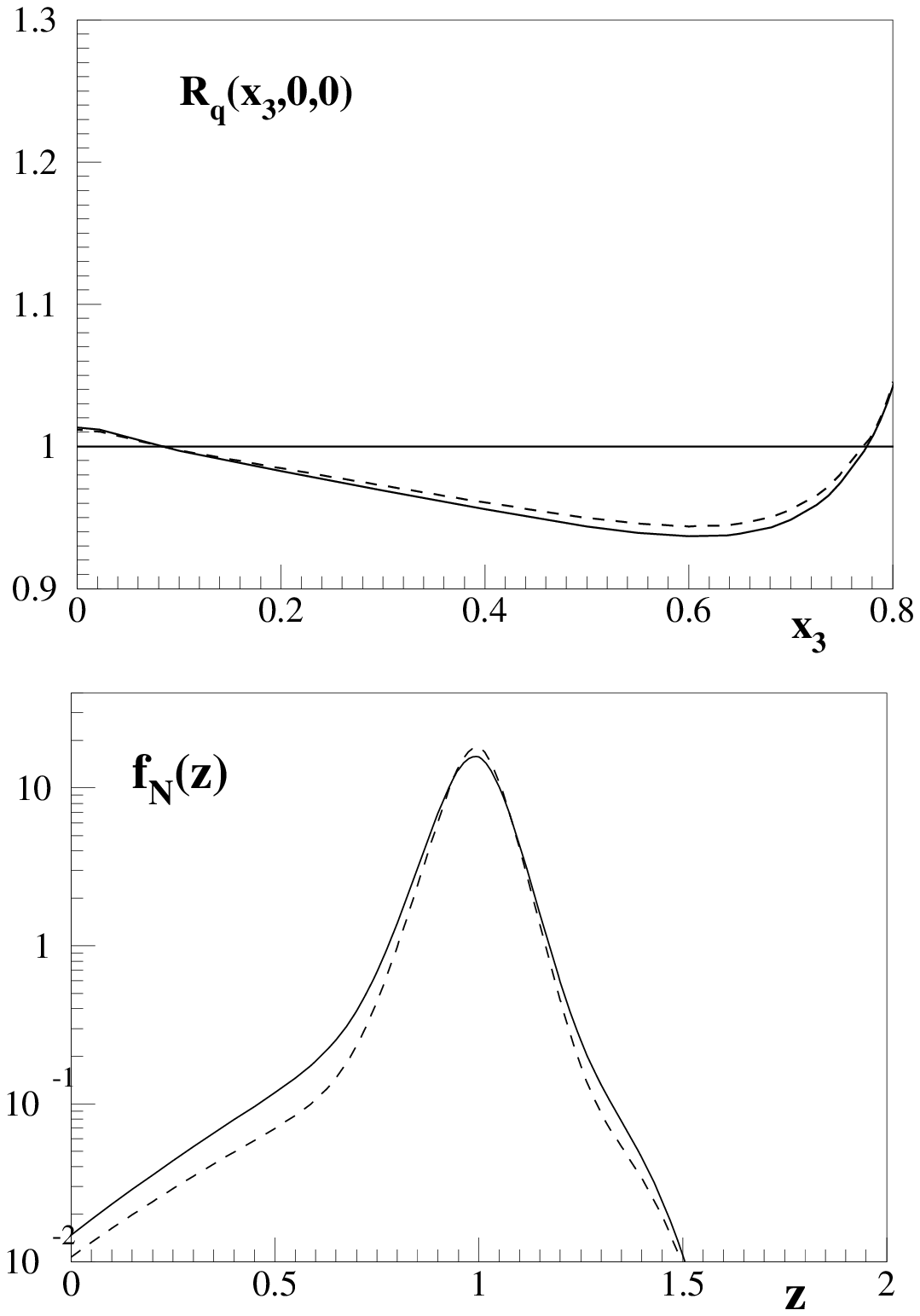}
\includegraphics{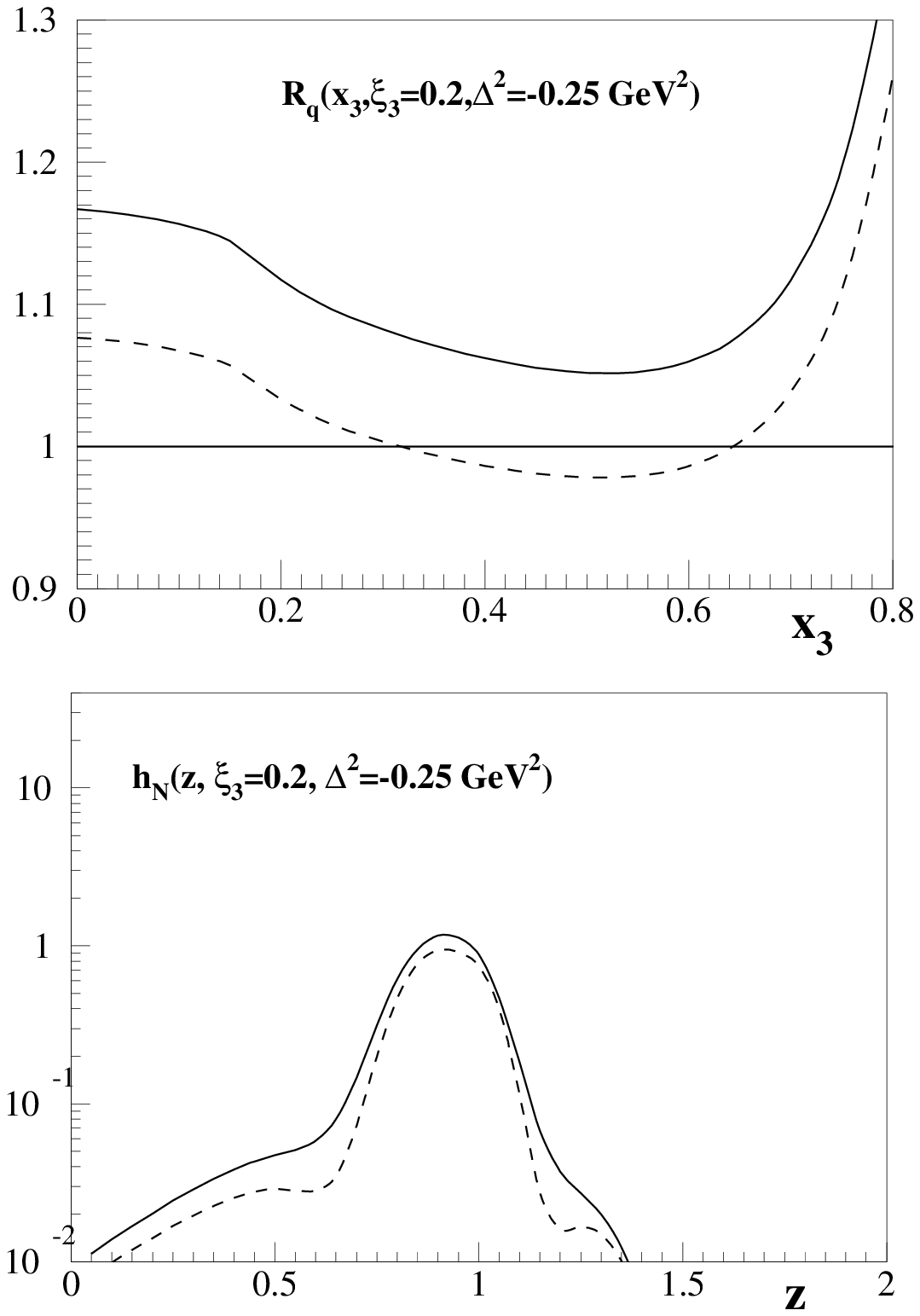}
\caption{
Left:
upper panel: the dashed (full) line represents
the ratio Eq. (\ref{rnew}), for the $u$ ($d$)
flavor, in the forward limit;
lower panel: the dashed (full) line represents
the light cone momentum distribution, Eq. (\ref{hq0f}),
for the proton (neutron) in $^3$He.
Right: the same as in the left panel, but at
$\Delta^2=-0.25$ GeV$^2$ and $\xi_3=0.2$.
}
\end{figure}

In Figs. 1 to 3, results will be presented concerning: 
A) flavor dependence of nuclear effects; B) binding effects; 
C) dependence on the nucleon-nucleon potential.
\\
A) Flavor dependence of nuclear effects.
In the upper left panel of
Fig. 1, the ratio Eq. (\ref{rnew}) is shown for the $u$ and $d$
flavor, in the forward limit,
as a function of $x_3=3x$. The trend is clearly EMC-like.
It is seen that nuclear effects for the $d$ flavour are very slightly bigger
than those for the $u$ flavour.
The reason is understood thinking that, in the forward limit,
the nuclear effects are governed by the light cone momentum distribution,
Eq. (\ref{hq0f}): no effects would be
found if such a function were a delta function, while effects get bigger
and bigger if its width increases.
In another panel of the same figure,
the light cone momentum distribution, Eq. (\ref{hq0f}),
for the proton (neutron) in $^3He$ is represented by the 
dashed  (full) line.
The neutron distribution is slightly wider than 
the proton one, meaning that the average momentum of the neutron
in $^3He$ is a little larger than the one of the proton. 
\begin{figure}[ht]
\vspace{4.7 cm}
\includegraphics{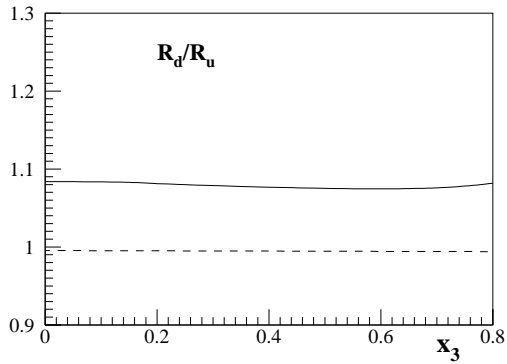}
\caption{
The ratio of the ratios Eq. (\ref{rnew}), for the $d$ to the
$u$ flavor, at  $\Delta^2=-0.25$ GeV$^2$ and $\xi_3=0.2$ (full line)
and in the forward limit (dashed line).
}
\end{figure}
Since the forward $d$ distribution
is more sensitive than the $u$ one to the neutron
light cone momentum distribution, nuclear effects for $d$ are slightly
larger than for $u$, as seen in the upper panel of the same figure.
In the same figure, the same analysis of Fig. 1 is performed,
but at $\Delta^2=-0.25$ GeV$^2$ and $\xi_3=3\xi=0.2$.
In this case, nuclear effects are governed by the non-diagonal
light cone momentum distribution, Eq. (\ref{hq0}), shown 
in the lower panel of the figure. In this case, the difference
between the neutron and proton distributions is quite bigger
than in the forward case, governing the difference 
in the ratio Eq. (\ref{rnew}) for the two flavors, which is of the order
of 10 \%, as it is seen in Fig. 3.
From Figs. 1-3 three main conclusions can be drawn.
1) if one infers properties of nuclear GPDs thinking
to those of nuclear PDs, conventional nuclear
effects as big as 10 \% can be easily
lost, or mistaken for exotic ones.
2) Secondly, this behavior is a typical conventional effect, being
a prediction of IA in DIS off nuclei. If a 10 \% effect would
be observable in experimental studies of nuclear GPDs,
the presence of such a flavor dependence, or its absence,
would be clear signatures of the reaction mechanism of DIS
off nuclei. Its presence would mean that the reaction involves 
essentially partons
inside nucleons, whose dynamics is governed by a realistic
potential in a conventional scenario; 
on the contrary, its absence would mean that,
in a different, exotic scenario,
other degrees of freedom have to be advocated. 
3) Eventually, it is clear that, for this kind of studies, $^3He$
is a unique target, for which experiments are worth to be done:
the flavor dependence cannot be investigated with
isoscalar targets, such as $^2H$ or $^4He$, while for heavier nuclei
calculations cannot be performed with comparable precision.
\begin{figure}[ht]
\vspace{9.0cm}
\includegraphics{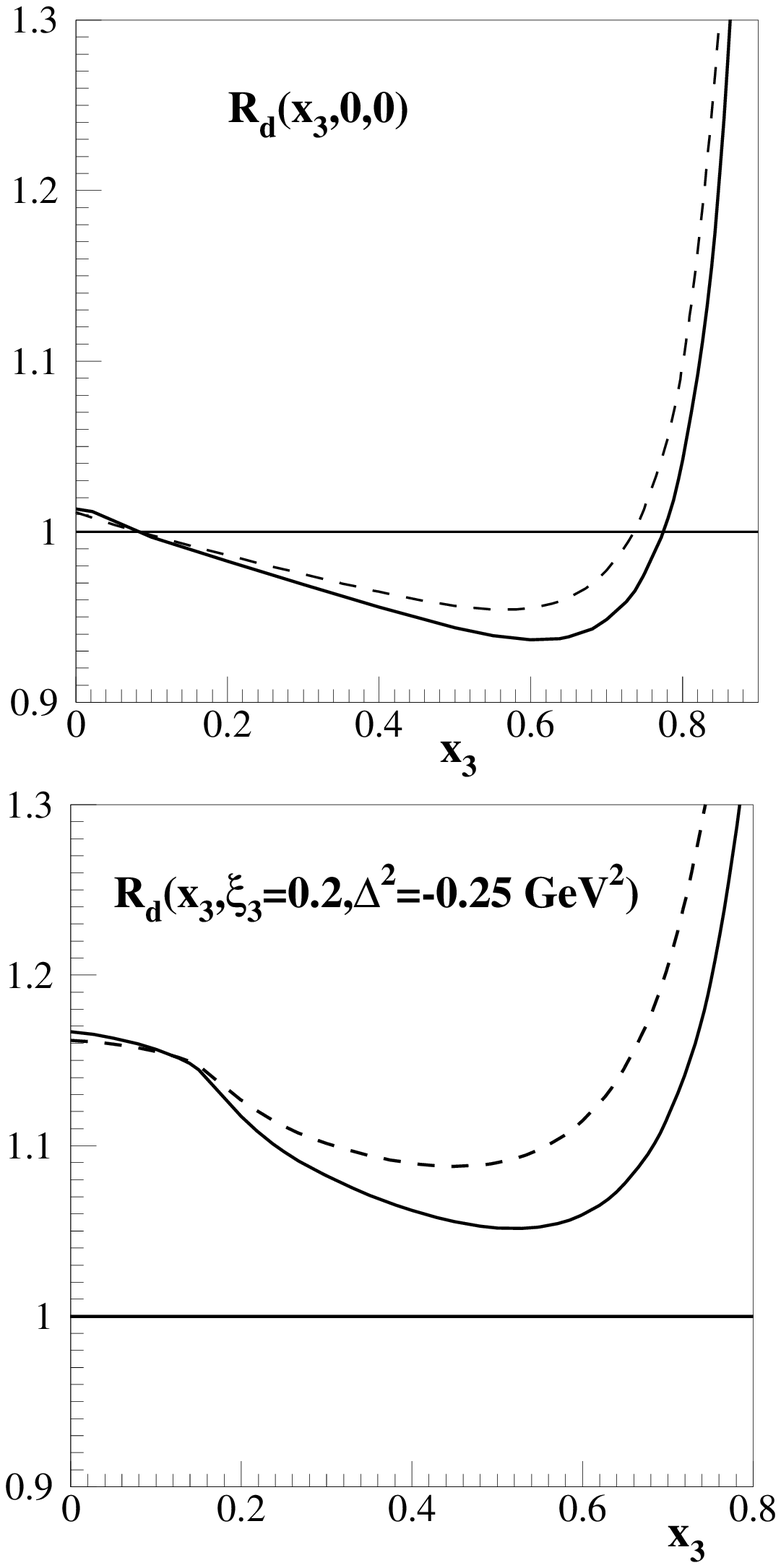}
\includegraphics{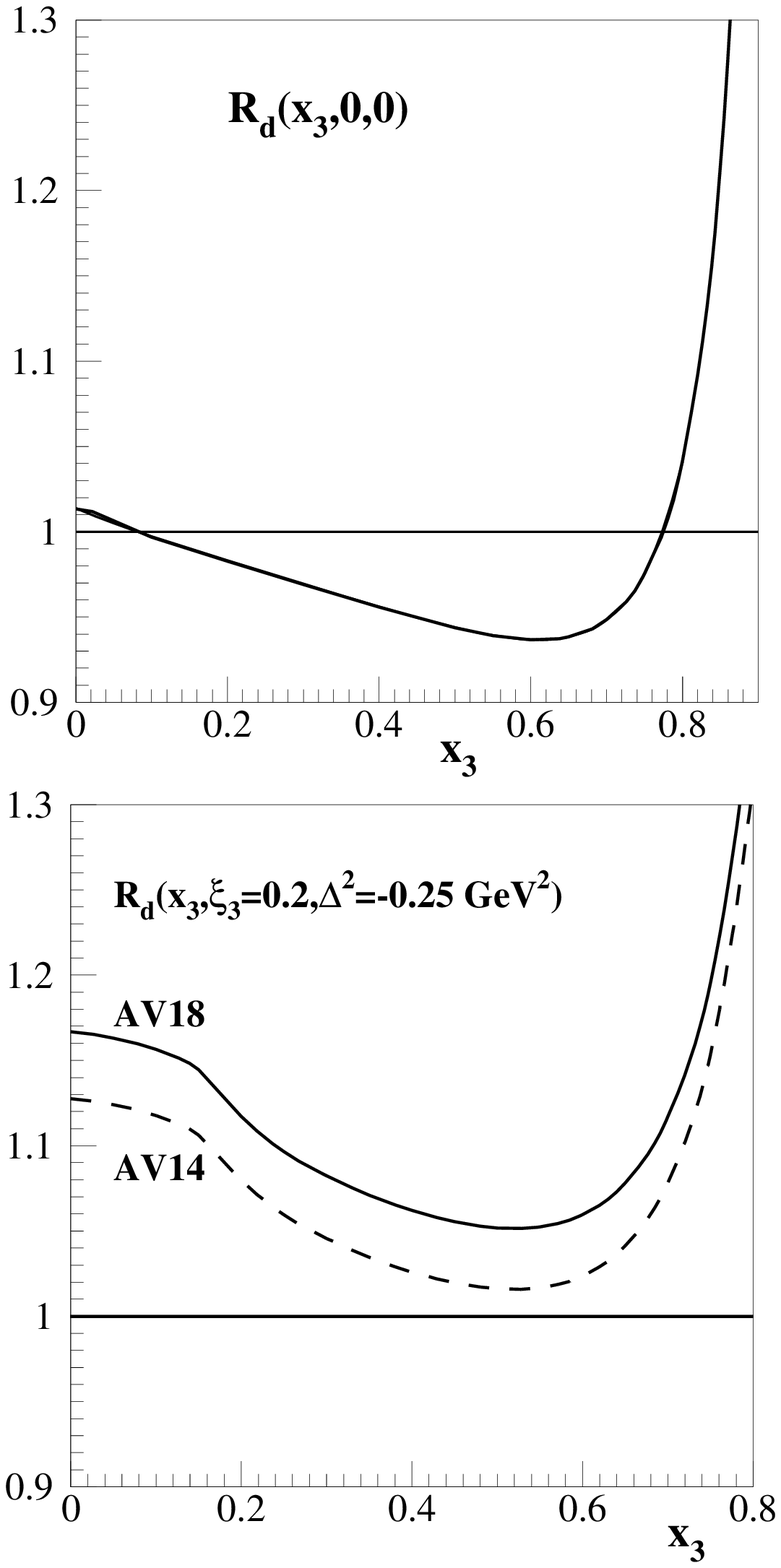}
\caption{
Left: upper panel: the ratio Eq. (\ref{rnew}), in the forward limit,
for the $d$ flavor,
corresponding to the full result of the present approach
(full line), compared with the one obtained using in the
numerator the approximation Eq. (\ref{clos}) with $\bar E^*=0$,
i.e., using a momentum distribution instead of a spectral function
(dashed line);
lower panel: the same as before, but 
evaluated at $\Delta^2=-0.25$ GeV$^2$ and $\xi_3=0.2$.  
Right: upper panel: the ratio Eq. (\ref{rnew}), in the forward
limit, for the $d$ flavor,
corresponding to the full result of the present approach,
where use is made of the AV18 interaction
(full line), compared with the one obtained using in the
numerator the AV14 interaction (dashed line): the two curves
cannot be distinguished;
lower panel: the same, but 
evaluated at $\Delta^2=-0.25$ GeV$^2$ and $\xi_3=0.2$: now the curves are
distinguishable.  
}
\end{figure}
\\
B) Binding effects.
In the previous section it has been explained how
Eq. (\ref{main}) takes
into account properly the nucleon momentum and energy distributions
through a non-diagonal spectral function.
In the following,
the result obtained neglecting binding effects, i.e.,
by using a momentum distribution
instead of a spectral function, will be shown.
When a momentum distribution is used instead of a spectral
function, not only the IA, but also another approximation,
the so called ``closure approximation'', has been used:
an average excitation energy, $\bar E^*$, has been inserted
in the expression of the delta function appearing in the definition
of the spectral function Eq. (\ref{spectral}), so that the completeness
of the two body recoiling states can be used \cite{cio}:
\be
P_N^3(\vec p, \vec p + \vec \Delta, E)  & \simeq &
\, \bar{\sum}_{M} 
\sum_{s}
\bra \vec P'M | a_{\vec p + \vec \Delta,s}
a^\dagger_{\vec p, s}| \vec P M \ket
\delta(E - E_{min} - \bar E^*)
\nonumber
\\
& = &
n(\vec p, \vec p + \vec \Delta)\,\delta(E - E_{min} - \bar E^*)~,
\label{clos}
\eq
and the spectral function is approximated by a
one-body non diagonal momentum distribution times
a delta function defining an average 
value of the removal energy.
Whenever the momentum distribution is used instead of the spectral
function, in addition to the IA the above closure approximation has been
used assuming  $\bar E^*=0$, i.e., binding effects have been completely
neglected.
The difference between the full calculation and the
one using the momentum distribution, for the ratio
Eq. (\ref{rnew}), is shown in Fig. 3.
It is seen that, while the difference is a few percent
in the forward limit, it grows in the non-forward case,
becoming an effect of 5 \% to 10 \% between 
$x = 0.4$ and $0.7$.
From this analysis, the same three main conclusions,
arisen in the study of the flavor dependence can be drawn.
\\
C) Dependence on the nucleon-nucleon potential.
In Fig. 3, the difference is shown between the full calculation,
Eq. (\ref{rnew}), evaluated with the AV18 interaction \cite{av18}, 
and the same quantity, evaluated by means of the AV14 one.
It is seen that there is basically no difference in the forward
limit, confirming previous findings in inclusive DIS
\cite{io2}, while a sizable difference is seen in the non forward case
(preliminary results of this behavior have been accounted for in a talk
at a Conference \cite{io}).
From these analyses the same conclusions of the previous two subsections
can be drawn.
We note on passing that a difference between observables evaluated
using AV18 and AV14 potentials is not easily found, in particular
in inclusive DIS.

\section{GPDs for the $^3H$ target}
\begin{figure}[ht]
\vspace{4.7 cm}
\includegraphics{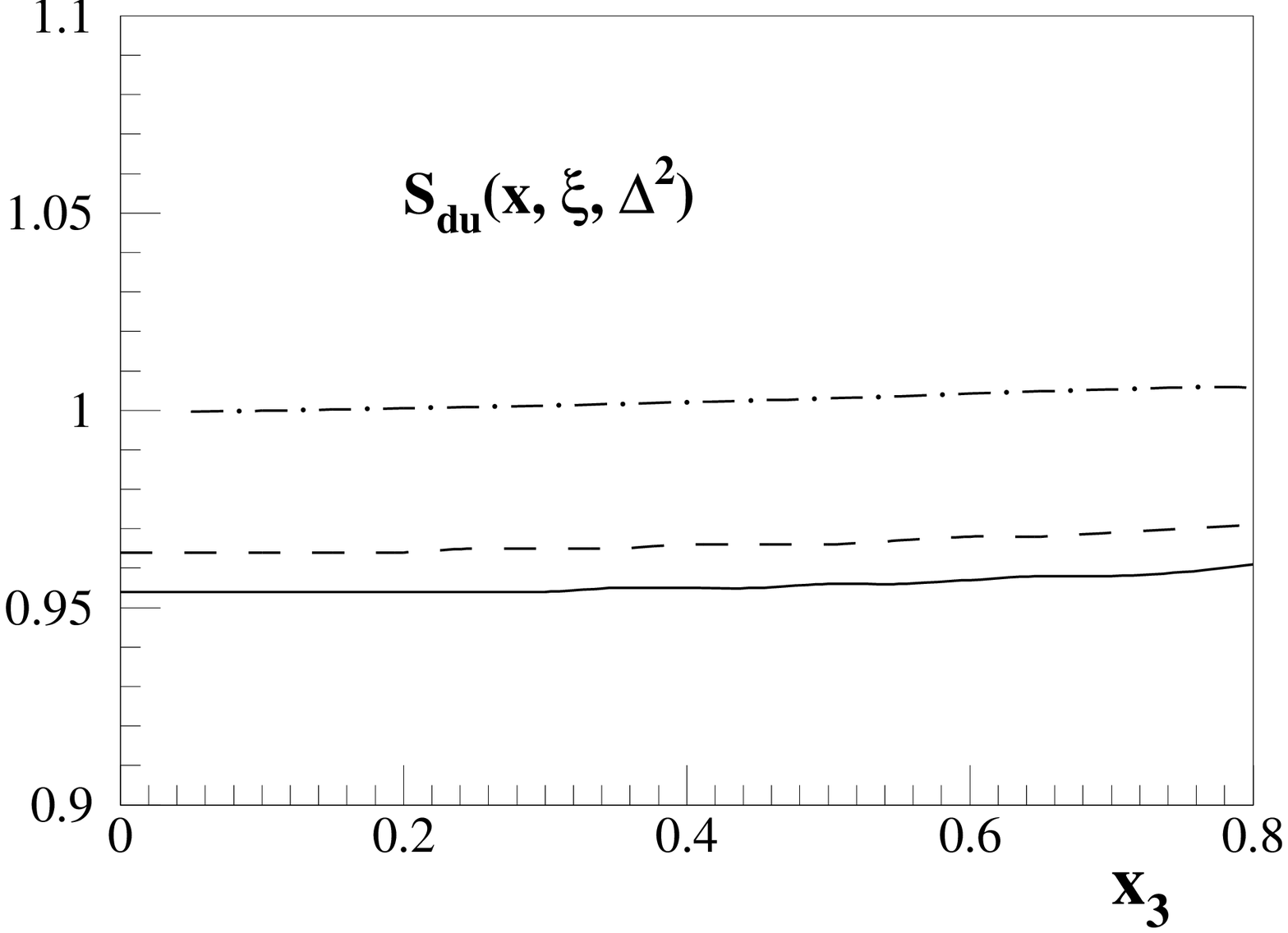}
\caption{
The ratio Eq. (\ref{extr}), in the forward limit (dot-dashed line),
at $\Delta^2=-0.15$ GeV$^2$ and $\xi_3=0.1$ (dashed line),
and at
$\Delta^2=-0.25$ GeV$^2$ and $\xi_3=0.2$ (full line).
}
\end{figure}
In the perspective of using $^3H$ targets after the 12 GeV upgrade
of JLab \cite{wp}, it is useful to address what could be learnt 
from simultaneous measurements with trinucleon targets, $^3He$ and $^3H$.
The procedure proposed firstly in Ref. \cite{thom1} for the unpolarized
DIS to extract, with unprecedented precision,
the ratio of down to up quarks in the proton, $d(x)/u(x)$,
at large Bjorken $x$, 
is extended here to the case of the GPDs of trinucleons.
To minimize nuclear effects, 
the following ``super-ratio'', a generalization
of the one proposed in Ref. \cite{thom1},  can be defined
\be
S_{qq'}(x,\xi,\Delta^2) = R_q^H(x,\xi,\Delta^2)/R_{q'}^T(x,\xi,\Delta^2)~,
\label{sr}
\eq
where the ratio
\be
R_q^A(x,\xi,\Delta^2)= { H_q^A(x,\xi,\Delta^2)
\over
Z_A H_q^p(x,\xi,\Delta^2) + N_A H_q^n(x,\xi,\Delta^2) }~,
\label{rsr}
\eq
has been introduced for $^3He$ ($A=H$) and $^3H$ ($A=T$),
with $q=u,d$, $Z_A (N_A)$ the number of protons (neutrons)
in the nucleus $A$, and $H_q^N(x,\xi,\Delta^2)$ the GPD
of the quark $q$ in the nucleon $N=p,n$.
Now, using the isospin symmetry of GPDs,
we can call
\begin{eqnarray}
H_u(x,\xi,\Delta^2) & = &
H_u^p(x,\xi,\Delta^2)
=H_d^n(x,\xi,\Delta^2)~,
\\
H_d(x,\xi,\Delta^2) & = &
H_d^p(x,\xi,\Delta^2)
=H_u^n(x,\xi,\Delta^2)~,
\end{eqnarray}
so that Eq. (\ref{sr}) is given, for example for $q=d$ and $q'=u$,
by the simple relation
\bq
S_{du} (x,\xi,\Delta^2)= { H_d^H(x,\xi,\Delta^2) \over H_u^T(x,\xi,\Delta^2)}~,
\label{extr}
\eq
a quantity in principle observable.
In the IA approach discussed here, using Eq. (\ref{main})
to calculate the nuclear GPDs, one has therefore
\bq
S_{du} (x,\xi,\Delta^2)= { 
\int_x^1 {dz \over z} \left \{ 
h_p^H(z,\xi,\Delta^2) H_d \left({x \over z},{\xi \over z },\Delta^2 \right )
+
h_n^H(z,\xi,\Delta^2) H_u \left({x \over z},{\xi \over z },\Delta^2 \right )
\right \}
\over 
\int_x^1 {dz \over z} \left \{ 
h_n^T(z,\xi,\Delta^2) H_d \left({x \over z},{\xi \over z },\Delta^2 \right )
+
h_p^T(z,\xi,\Delta^2) H_u \left({x \over z},{\xi \over z },\Delta^2 \right )
\right \} }~,
\label{extr}
\eq
where $h_{p(n)}^{H(T)}(z,\xi,\Delta^2)$ represents 
the light cone off diagonal momentum distribution for
the proton (neutron) in $^3He$ ($^3H$).
If the Isospin Symmetry were valid at the nuclear level,
one should have 
$h_{p}^{H}(z,\xi,\Delta^2) = h_{n}^{T}(z,\xi,\Delta^2)$, 
and
$h_{n}^{H}(z,\xi,\Delta^2) = h_{p}^{T}(z,\xi,\Delta^2)$,
so that the ratio Eq. (\ref{extr}) would be identically 1.
From the previous analysis,
it is clear anyway
that these relations are only approximately true,
and some deviations are expected.
In Fig. 4, the super-ratio $S_{du} (x,\xi,\Delta^2)$, Eq. (\ref{sr}),
evaluated by using the AV18 interaction for the nuclear GPDs
in Eq. (\ref{main}),
taking into account therefore the Coulomb interaction between
the protons in $^3He$ and a weak charge independence breaking term, 
is shown for different values of $\Delta^2 \leq 0.25$ GeV$^2$ and $\xi$.
While it is seen that, as expected, $S_{du} (x,\xi,\Delta^2)$
is not exactly 1 and the difference gets bigger with increasing
$\Delta^2$ and $\xi$, for the low values of $\Delta^2$ and $\xi$ relevant
for the present investigation of GPDs, 
such a difference keeps being a few percent one.
It would be very interesting to measure this ratio experimentally.
If strong deviations from this predicted behavior were observed,
there would be a clear evidence that the description in terms of
IA, i.e. in terms of the conventional scenario of partons confined
in nucleons bound together by a realistic interaction, breaks down.
In other words one could have a clear signature of possible interesting
exotic effects.

In summary,
conventional nuclear effects on
the unpolarized quark GPD for the trinucleons
have been described, using
a realistic microscopic calculation. 
The issue of applying the obtained GPDs to  
estimate cross-sections and to establish
the feasibility of experiments, is in progress
and will be presented elsewhere.

\end{document}